\documentclass[a4paper,11pt,reqno]{article}
\usepackage{amssymb}
\usepackage{amsmath}
\usepackage{latexsym}
\usepackage{amssymb,amsbsy,amsmath,amsfonts,amssymb,amscd,amsthm}
\usepackage{accents}	
\usepackage[T1]{fontenc}
\usepackage[utf8]{inputenc}
\usepackage{graphicx}
\graphicspath{{./}{Figure/}}
\usepackage{hyperref}
\usepackage{dsfont}
\usepackage{mathtools}
\usepackage{pdfsync}
\usepackage{epsfig,epstopdf}
\usepackage{caption}
\usepackage{xcolor}
\usepackage{hyperref}
\usepackage[title]{appendix}
\usepackage{subcaption}
\usepackage{float}
\usepackage{color}
\usepackage{verbatim}
\usepackage{appendix}
\usepackage[inline]{enumitem}
\newtheorem{thrm}[]{Theorem}

\newcommand{\beq}{\begin{equation}}
\newcommand{\eeq}{\end{equation}}
\newcommand{\beqa}{\begin{eqnarray}}
\newcommand{\eeqa}{\end{eqnarray}}

\newcommand{\di}{\, {\rm d}}
\newcommand {\e}  {\varepsilon}

\begin{document}
\title{Optimal control on a heterogeneous SI epidemic model} 
\author{Elisa Paparelli\thanks{Sorbonne Université, Université Paris Cité, CNRS, Laboratoire Jacques-Louis Lions, LJLL, F-75005 Paris, France}
}
\date{}
\maketitle
\begin{abstract}
This work addresses an optimal control problem for a SI epidemic model incorporating heterogeneities in resistance and viral load at the population level. Building upon the heterogeneous SI framework developed in \cite{tea23}, a minimization problem constrained to the macroscopic counterpart of the SI dynamics derived therein is proposed. Unlike traditional optimal control problems in homogeneous epidemic models, the present approach focuses on an optimal control problem that accounts for population heterogeneity, offering insights from a microscale perspective. The contribution aims to minimize the final size of the infection within a finite time horizon by developing a pharmaceutical strategy, under a supply constraint that translates into an integral equality constraint in the control function. By applying the Pontryagin Minimum Principle, a characterization of an optimal control is provided.
\end{abstract}
\noindent {\textbf{Keywords:} epidemic model, optimal control, pharmaceutical intervention, multiscale approach }
\section{Introduction}
\label{sec1}
In this work, we propose an optimal control problem based on a heterogeneous SI epidemic phenomenon. In particular, an epidemic model incorporating heterogeneity in resistance and viral load has been developed in \cite{tea23}, starting from an agent-based formulation that describes both the interaction between agents in different compartments and the natural evolution of resistance and viral load. Analytical results therein have shown that, in this scenario, the dynamics leads the whole population to be infected, or to the endemic scenario for a specific choice of the parameters involved. 

The literature on optimal control in infectious diseases is broad \cite{behncke2000,neilan2010,sethi1978,sharomi2017}, spanning both non-pharmaceutical and pharmaceutical interventions. Several studies have applied control strategies to classical SIR models \cite{lin2010,kruse2020}, while others examine pharmaceutical protocols such as vaccination \cite{cianfanelli2021,ledzewicz2011}, including combinations with quarantine and protective devices to minimize eradication time \cite{bolzoni2017}. Following the COVID-19 outbreak, most compartmental models initially focused on non-pharmaceutical interventions \cite{flaxman2020,gatto2020,kucharski2020,giordano2021}, including infectivity-rate reduction across age or geographic groups \cite{acemoglu2021,gondim2020,lu2021}. On the pharmaceutical side, several works model vaccination as a flow of individuals from the susceptible to the recovered compartment \cite{mandal2020,libotte2020}, and some studies combine multiple interventions, including pharmaceutical ones, within a single framework \cite{abbasi2020,asamoah2021,choi2021,mondal2022,shen2021}. 
A further strand of literature imposes additional state and/or control constraints, for instance to prevent hospital overcrowding \cite{kohler2021}, to account for limited resources \cite{bolzoni2019}, or other mixed state-control constraints \cite{biswas2014,caulkins2021,lauriere2020,pinho2015,miclo2022}.  \cite{caulkins2021,kantner2020} relate control to workforce and economic consequences, while \cite{aspri2021, paparelli24} considers a functional that more plausibly accounts for the impact of restrictions on workforce and gross domestic product, alongside the loss of human lives. 
More recently, a complementary line of research has addressed optimal control problems for epidemic dynamics in which population heterogeneity is treated as a social trait, leading to socially structured formulations where the control acts on the resulting heterogeneous interaction patterns \cite{albi2021control,albi2021lockdown,dimarco2022,franceschi2024}.

The individual-based model that drives the dynamics under analysis is summarized as follows.\\
Each agent is labelled by the triplet $(X_t, R_t, V_t)$ where, at each time $t$, $X$ describes the compartment (S - Susceptible, I - Infected) which the agent belongs to, and $R$ and $V\in [0,1]$, respectively the level of resistance and viral load. The evolution of  $(X_t, R_t, V_t)$ is driven by the following individual-based dynamics:
\beq
\label{eq:microdynlaws}
\begin{cases}
X_{t + \Delta t} = (1 - \Theta_X) \, X_t + \Theta_X \, X'_t,
\\
R_{t + \Delta t} = (1 - \Theta_R) \, R_t + \Theta_R \, R'_t,
\\
V_{t + \Delta t} = (1 - \Theta_X) \Big[(1 - \Theta_V) \, V_t + \Theta_V \, V'_t \Big] + \Theta_X \, V''_t,
\end{cases}
\eeq
where
\beq
\label{eq:Theta}
\Theta_{\alpha} \sim {\rm Bernoulli}\left(\theta_{\alpha} \, \Delta t \right) \; \text{ with } \; \theta_{\alpha} \in \mathbb{R}^*_+  \; \text{ for } \; \alpha \in \{X, R, V\},
\eeq
\beq
\label{eq:Rp}
\left(R'_t | X_t, R_t\right) \sim J_S(r'|x_i, r) \; \text{ with } \; J_S(r'|x_i, r) := K_S(r'| r)\, \delta_{x_i,S} + \delta_{r}(r')\, \delta_{x_i,I},
\eeq
and
\beq
\label{eq:Vp}
\left(V'_t | X_t, V_t\right) \sim J_I(v'|x_i, v) \; \text{ with } \; J_I(v'|x_i, v) := \delta_{v}(v')\,\delta_{x_i,S} +K_I(v'| v)\,\delta_{x_i,I},
\eeq
where $\delta_{b}(a)$ is the Dirac delta centred in $a=b$ and $\delta_{\beta, \alpha}$ is the Kronecker delta centred in $\alpha = \beta$. Moreover, we let
\beq
\label{eq:XpVpp}
\left(X'_t, V''_t | X_t, R_t, V_t, X^*_t, V^*_t \right) \sim T(x'_i, v'' | x_i, r, v, x^*_j, v^*) 
\eeq
with 
\beq
\label{eq:T}
T(x'_i, v'' | x_i, r, v, x^*_j, v^*) := 
\begin{cases}
q(r, v^*) \, Q(v'' |r, v^*) \, \delta_{I, x'_i} + \left(1 - q(r, v^*)\right) \, \delta_{v''}(v) \, \delta_{S, x'_i}, \;\;\;\;\;\,\quad &\text{if } x_i = S, \; x^*_j = I,
\\
\delta_{v}(v'') \, \delta_{S, x'_i}, \quad \quad \quad \quad \quad \quad \quad\quad\quad \quad \quad \quad \quad \quad \quad\quad\quad \quad  &\text{if } x_i = S, \; x^*_j = S,
\\
\delta_{v}(v'') \, \delta_{I, x'_i}, \quad \quad \quad \quad \quad \quad \quad\quad\quad \quad \quad \quad \quad \quad \quad\quad\quad \quad \; &\text{if } x_i = I. 
\end{cases}
\eeq
The probability distribution $K_S(r'|r)$ in \eqref{eq:Rp} represents the distribution of the new resistance level $r'$ starting from the level $r$, when the switch occurs in a susceptible agent. In analogous way, in \eqref{eq:Vp}, $K_I(v'|v)$ represents the transition probability to a new viral load level $v'$ in an infected agent with viral load $v$ when the switch occurs. In definition \eqref{eq:T} the function $q(r,v^*)$ is the rate at which the contagion after the interaction between a susceptible with resistance level $r$ and an infected with viral load level $v^*$ occurs. The function $Q(v''| r,v^*)$ indicates the distribution of the new viral load acquired after that kind of contagion. 
 Following the derivation of the \emph{mesoscopic} dynamics describing the distribution evolution of resistance in susceptibles and viral load in infected individuals, provided by a coupled system of Integro-Differential Equations, under appropriate assumptions, the dynamics of macroscopic quantities $N_S(t)$, $N_I(t)$, $P_S(t)$, and $P_I(t)$ is derived. The functions $N_S(t)$ and $N_I(t)$ represent the total number of susceptibles and infected individuals at time $t$, respectively, while $P_S(t)$ and $P_I(t)$ represent the first moment of resistance  and viral load, respectively in susceptibles and infected individuals at time $t$. Given the fact that $N_S+N_I \equiv 1$ at all times, the macroscopic counterpart of the heterogeneous SI epidemic model can be written as a system of ODEs for quantities $N_I$, $P_S$, and $P_I$ as follows: 
\beq \label{controlled:odeNiPsPi}
\begin{cases}
\displaystyle{\dot N_I=P_I \Bigl[q(0,1)(1-N_I-P_S)+q(1,1)P_S \Bigr]},
\\\\
\displaystyle{\dot P_S =-q(1,1)\,P_I\,P_S},
\\\\
\displaystyle{\dot P_I=P_I \Bigl [ \overline Q(0,1)\,q(0,1)\,(1-N_I-P_S)+\overline Q(1,1)\, q(1,1)\,P_S \Bigr ] }, 
\end{cases}
\eeq
where $q(0,1),q(1,1),\overline Q(0,1), \overline Q(1,1) \in [0,1]$. Since it is assumed that the support of the variables tracking resistance and viral load is the compact set $[0,1]$, the parameter $q(0,1)$ indicates the probability that a susceptible with the lowest resistance, $0$, gets the infection after a contact with an infected with highest level of viral load, $1$, and $\overline Q(0,1)$ is the mean viral load acquired by the susceptible after that contagion. The parameter $q(1,1)$ indicates the probability that a susceptible with the highest level of resistance, $1$, gets the infection after a contact with an infected with highest level of viral load, $1$ and $\overline Q(1,1)$ is the mean viral load acquired by the susceptible after that contagion. Starting from the SI model that incorporates the heterogeneities on resistance and viral load levels recalled above, we introduce a minimization problem constrained to the macroscopic counterpart equations that captures the heterogeneity structure as well. Compared to optimal control problems in epidemics formulated as homogenous compartmental systems, our challenge is to formulate and study an optimal control problem capturing the heterogeneity structure on population, providing meaningful insight under the \emph{microscale} point of view. 
The contribution focuses on analyzing an optimal control problem aimed at minimizing the final size of the infection within a finite time horizon by controlling the quantity $\overline Q(0,1)$. 
For further details about the derivation of the dynamics \eqref{controlled:odeNiPsPi} and the asymptotic behavior analysis, see \cite{tea23}. \par 
In Section \ref{sec2} the optimal control problem is formulated; in Section \ref{sec3} the analytical results of the problem by characterizing the optimal control is established, and a sample of numerical results are provided. 
\section{Formulation of the optimal control problem} \label{sec2}
The first challenge is to formulate an optimal control problem constrained to the evolution of the macroscopic quantities 
that would provide the possibilities to control the multi-agent dynamics under the microscale point of view. We choose to contain the infection, by developing a pharmaceutical strategy that targets the population with the more propensity to be infected, i.e. on susceptible agents with the lowest level of resistance, which are also more easily identifiable under the medical point of view. 
Specifically, we focus on a scenario where: 
\begin{itemize}
\item[(i)] the vaccination schedule is expressed by the quantity $\overline Q(0,1)$, hence we define the time evolving control function $u(t):=\overline Q(t,0,1)$; in the individual-based dynamics \eqref{eq:microdynlaws}-\eqref{eq:T}, this assumption leads the distribution function $Q(t,v''|r,v^*)$ to be time-dependent;
\item[(ii)] we impose the control $u(t)$ to be limited, hence
\beq \label{ass:u}
u_{min}\le u(t) \le u_{max},
\eeq
with $0\le u_{min} \le u_{max}\le 1$. Moreover, it is natural to add a constraint on the intervention cost in the time horizon. This idea translates into the $L^1$-type constraint on the control function, i.e
\beq \label{ass:L1}
\int_0^T (u_{max}-u(t) )\, \di t =C, 
\eeq
where 
\beq \label{ass:C}
C=\tilde T \,(u_{max}-u_{min}),
\eeq
 with $\tilde T$ the sum of time intervals where the control could assume its minimum value and $0<\tilde T<T$.
This constraint symbolizes practical limitations in vaccine administration efforts, emphasizing the need to make full use of available resources; 
\item[(iii)] the goal is to define a strategy to control the system minimizing the final size of the infection, thus, the objective to be minimized is defined as
\beq \label{ass:obj}
J(u)=N_I(T),
\eeq
where $T>0$ is fixed and finite; 
\item[(iv)]  the initial conditions are fixed, thus
\beq \label{ass:ic}
N_{I}(0)\equiv N_{I,0}, \quad P_S(0) \equiv P_{S,0}, \quad P_I(0) \equiv P_{I,0}, \quad N_{I,0},  P_{S,0}, P_{I,0} \in (0,1).
\eeq
\end{itemize}
Notice that, the $L^1$ constraint \eqref{ass:L1} is equivalent to add a new variable defined as $\displaystyle x(t):=\int_0^t (u_{max}-u(\tau) )\, \di \tau$, together with the boundary condition
\beq \label{ass:xb}
x(0)=0, \quad x(T)=C.
\eeq
Hence, from \eqref{controlled:odeNiPsPi}, we obtain the ODEs system  
\beq \label{controlled:odeNiPsPix}
\begin{cases}
\displaystyle{\dot N_I=P_I \Bigl[q(0,1)(1-N_I-P_S)+q(1,1)P_S\Bigr]},
\\\\
\displaystyle{\dot P_S =-q(1,1)\,P_I\,P_S},
\\\\
\displaystyle{\dot P_I=P_I \Bigl [ u(t)\,q(0,1)\,(1-N_I-P_S)+\overline Q(1,1)\, q(1,1)\,P_S\Bigr]},
\\\\
\displaystyle{\dot x=u_{max}-u(t)}.
\end{cases}
\eeq
The optimal control problem is formulated as follows:
\begin{equation}\label{ocp2}
\left. \text{(OCP)     }
\begin{aligned}
&\min_{u \in \Omega} J(u)\\
&\text{ subject to \eqref{controlled:odeNiPsPix} and boundary conditions \eqref{ass:ic}, \eqref{ass:xb}}
\end{aligned}
\right \rbrace
\end{equation}
where $J(u)$ is defined via \eqref{ass:obj} and the control set as
\beq \label{ass:omega}
\Omega=\{ u \in L^1 (0,T), u_{min}\le u(t)\le u_{max}  \},
\eeq
with $T>0$ fixed, $0\le u_{min}\le u_{max}\le 1$, $C>0$ according to \eqref{ass:C}.

\section{Main results} \label{sec3}
By standard arguments, the existence of the solution of the optimal control problem (OCP) is established (see for example \cite{flemingrishel1975}, Theorem 2.1, pag. 63). 
Making use of the necessary conditions provided by the Pontryagin Minimum Principle, our main results deal with the characterization of the optimal control for a sufficiently small time horizon $T$.
\paragraph{Pontryagin Minimum Principle}The assurance of an optimal solution's existence ensures that the optimal solution meets the necessary conditions provided by  Pontryagin Principle, see for example \cite{hartl1995}. Letting $(u, N_I, P_S, P_I, x)$ the optimal pair control-state of the minimization problem \eqref{ocp2}, then there exist absolutely continuous functions $(\lambda_1, \lambda_2, \lambda_3, \lambda_4)$ and the constants $\lambda_0 \ge 0$ and $\beta \in \mathbb R$ such that $(\lambda_0, \lambda_1, \lambda_2, \lambda_3, \lambda_4,\beta) \neq 0$ for every $t \in [0,T]$ and the following conditions hold:
\beq \label{opt}
u(t)= \arg\min_{u \in \Omega} \mathcal H(N_I, P_S, P_I,x, u, \lambda_1, \lambda_2, \lambda_3, \lambda_4),
\eeq
where $\mathcal H$ is the Hamiltonian of the system, defined as
\beq \label{ham}
\begin{aligned} 
\mathcal H=&\lambda_1\, \Bigl \{ P_I \Big[ q(0,1)\,(1-N_I-P_S)+q(1,1)\,P_S) \Big] \Bigr\} + \lambda_2\, \Bigl \{ -q(1,1)\,P_I\,P_S     \Bigr\}+\\
&\lambda_3 \, \Bigl \{   P_I \Big [ u(t)\,q(0,1)\,(1-N_I-P_S)+\overline Q(1,1)\, q(1,1)\,P_S \Big ]         \Bigr \} + \lambda_4\, \Bigl \{  u_{max}-u(t)   \Bigr \},
\end{aligned}
\eeq
and the evolution of the adjoint functions $\lambda_1$, $\lambda_2$, $\lambda_3$ and $\lambda_4$ is driven by the ODEs system
\beq \label{eq:adj}
\resizebox{.93\textwidth}{!}{$
\begin{cases}
\displaystyle{\dot \lambda_1=-\frac{\di \mathcal H}{\di N_I}= q(0,1)\,P_I\, \lambda_1 + q(0,1)\,u(t)\,P_I\,\lambda_3 },
\\\\
\displaystyle{\dot \lambda_2 = -\frac{\di \mathcal H}{\di P_S}= [q(0,1)-q(1,1)]\,P_I\,\lambda_1 + q(1,1)\,P_I\,\lambda_2 + [q(0,1)\,u(t)-q(1,1)\,\overline Q(1,1) ]\,P_I\,\lambda_3   },
\\\\
\displaystyle{\dot \lambda_3 =-\frac{\di \mathcal H}{\di P_I}= [ -q(0,1)\,(1-N_I-P_S)-q(1,1)\,P_S] \lambda_1+q(1,1)\,P_S\,\lambda_2+-[q(0,1)\,u(t)\,(1-N_I-P_S)+q(1,1)\,\overline Q(1,1)\,P_S ]\, \lambda_3}
\\\\
\displaystyle{\dot \lambda_4=-\frac{\di \mathcal H}{\di x}= 0 },
\end{cases}$}
\eeq
complemented by the boundary conditions 
\beq \label{adja} 
\lambda_1(T)=\lambda_0 \ge 0,\quad \lambda_4(T)=\beta,\quad \lambda_2(T)=\lambda_3(T)=0.
\eeq
Since the linear dependence of the Hamiltonian \eqref{ham} with respect to the control, the optimality condition \eqref{opt} leads to
\beq\label{optc}
u(t)= 
\begin{dcases}
 u_{max} \quad &\text{ if }\,\, \phi(t)<0,
\\
u^* \quad \quad &\text{  if }\,\, \phi(t)=0,
\\
u_{min} \quad &\text{   if }\, \,\phi(t)>0,
\end{dcases}
\eeq
where $u^* \in (u_{min},u_{max})$ and $\phi(t)$ is the so-called switch function, defined as $\displaystyle \phi(t)\coloneqq \frac{ \di \mathcal H}{\di u}$,
indeed
\beq \label{def:switch}
\phi(t)=\lambda_3\,P_I\,q(0,1)\,(1-N_I-P_S)-\lambda_4.
\eeq
We can not exclude a priori the existence of intervals $I \subset [0,T]$ such that $\phi(t)=0$ for every $t\in I$, leading to singular controls within this interval. 
Before examining the optimal control, we aim to rule out the possibility of the control being abnormal, i.e. $\lambda_0=0$. We start by assuring that $\lambda_0 \neq 0$. We firstly notice that, from equation \eqref{eq:adj}$_4$, and from boundary conditions \eqref{adja}, 
\beq \nonumber
\lambda_4 (t) \equiv \beta \quad \text{for every } t \in [0,T].
\eeq
By absurd, we assume that $\lambda_0=0$. From \eqref{adja}, It would follow that $\lambda_1(T)=\lambda_2(T)=\lambda_3(T)=0$. Since the adjoint system \eqref{eq:adj} is linear with respect to $\lambda_1$, $\lambda_2$, $\lambda_3$, for existence and uniqueness of the solution, $\lambda_1(t)=\lambda_2(t)=\lambda_3(t)=0$ for every $t \in [0,T]$. This solution would lead the Hamiltonian \eqref{ham} to 
\beq \nonumber
\mathcal H=\lambda_4\,(u_{max}-u(t)) \quad \text{for every } t\in[0,T].
\eeq
From equation \eqref{eq:adj}$_4$, $\lambda_4$ is constant, and  if $\lambda_4 \neq 0$ from \eqref{optc}-\eqref{def:switch}, it would imply that $\phi$ has constant sign for every $t \in [0,T]$, hence $u(t)=u_{max}$ or $u(t)=u_{min}$ for every $t \in [0,T]$, that are not admissible controls for the $L^1$ constraint defined via \eqref{ass:L1}. On the other hand, if $\lambda_4=0$ for every $t \in [0,T]$, we get $(\lambda_0, \lambda_1, \lambda_2, \lambda_3, \lambda_4)\equiv 0$ in $[0,T]$, that is an absurd by the statement of the Pontryagin Principle. Hence $\lambda_0 >0$ and, without loss of generality, it can be normalized such that $\lambda_0=1$.\par
Now, we establish our main results regarding the analytical characterization of the optimal control in two different SI case settings. Indeed, we recall that the analytical results obtained  in \cite{tea23}, show different epidemic scenarios assuming $q(1,1)=0$ or $q(1,1) \neq 0$. The proofs are provided in the appendix.
\begin{thrm}\label{th1}
Let the assumption $q(1,1)=0$ hold. If the following condition is satisfied
\beq \label{condition:l1l3}
\Bigr(e^{u_{max}\,q(0,1)\, \overline B\, T}-1\Bigl)\,\Bigr(e^{q(0,1)\, \overline P\,T}-1\Bigl)\le1,
\eeq
where $\overline B=1-N_{I,0}-P_{S,0}$ and $\overline P=P_{I,0}+u_{max}\,(1-N_{I,0}-P_{S,0})$,
then an optimal control of the OCP \eqref{ocp2} is such that 
\beq \label{def:u_opt}
u(t)=
\begin{dcases}
u_{min} \quad &\text{ for } t \in [0, \tilde T) ,
\\
u_{max} \quad &\text{ for } t \in (\tilde T,T].
\end{dcases}
\eeq
\end{thrm}
\begin{thrm}\label{th2}
Let the assumptions $0< q(1,1)<q(0,1) \le 1$ and $0<\overline Q(1,1)\le1$ hold. If the following conditions are satisfied
\vspace{0.2cm}
\beq \label{prop43:cond1}
\frac{\overline G+P_{S,0}\,\Bigl [q(0,1)-q(1,1) \Bigr] \Bigl(1-e^{-q(1,1)\overline P T} \Bigr )}{\overline J-\frac{P_{S,0}}{\overline P}\tilde F \Bigl(1-e^{-q(1,1) \overline P T} \Bigr)\Bigl(e^{\overline JT}-1 \Bigr)}\Bigl(e^{\overline J T}-1 \Bigr) \le \frac{\Bigr [q(0,1)-q(1,1) \Bigl]\,P_{I,0}\,e^{-q(0,1)\overline PT}}{\overline F+\Bigr[q(0,1)-q(1,1) \Bigl]\,P_{I,0}\,u_{max}\,\Bigl(1-e^{-q(0,1)\overline P T} \Bigr)},
\eeq
\beq \label{sign:l3}
\overline J- \frac{P_{S,0}}{\overline P} \tilde F \Bigr(1-e^{-q(1,1) \overline P T} \Bigl ) \Bigr (e^{\overline J T}-1 \Bigl)>0,
\eeq
where 
\begin{equation*}
\overline J=q(0,1)\,u_{max}\,(1-N_{I,0})+q(1,1)\,\overline Q(1,1)\, P_{S,0}, \quad \overline G=q(0,1)\,(1-N_{I,0})+q(1,1)\,P_{S,0},
\end{equation*}
\begin{equation*}
\overline P=P_{I,0}+u_{max}\,(1-N_{I,0}-P_{S,0})+\overline Q(1,1)P_{S,0},\quad \tilde F=\Bigl [q(0,1)\,u_{max}-q(1,1)\,\overline Q(1,1) \Bigl] \, \overline P,
\end{equation*}
\begin{equation*}
\overline F=-\Bigl [q(0,1)\,u_{min}-q(1,1)\,\overline Q(1,1) \Bigl] \, \overline P,\\
\end{equation*}
then an optimal control of the OCP \eqref{ocp2} is characterized by \eqref{def:u_opt}.
\end{thrm}
A sample of numerical results, meeting the assumptions of the aforementioned theorems, is displayed below. The figures show consistency between the numerical solutions and the analytical results previously established. They also illustrate the agreement between the macroscopic dynamics and the individual-based model dynamics when the optimal control is reinserted.
\begin{figure}[H]
\centering
\subcaptionbox{\label{SIocp_q11=0}}
{\includegraphics[trim={0cm 7cm 0cm 7cm}, clip, width=.45\textwidth, height=.35\textwidth]{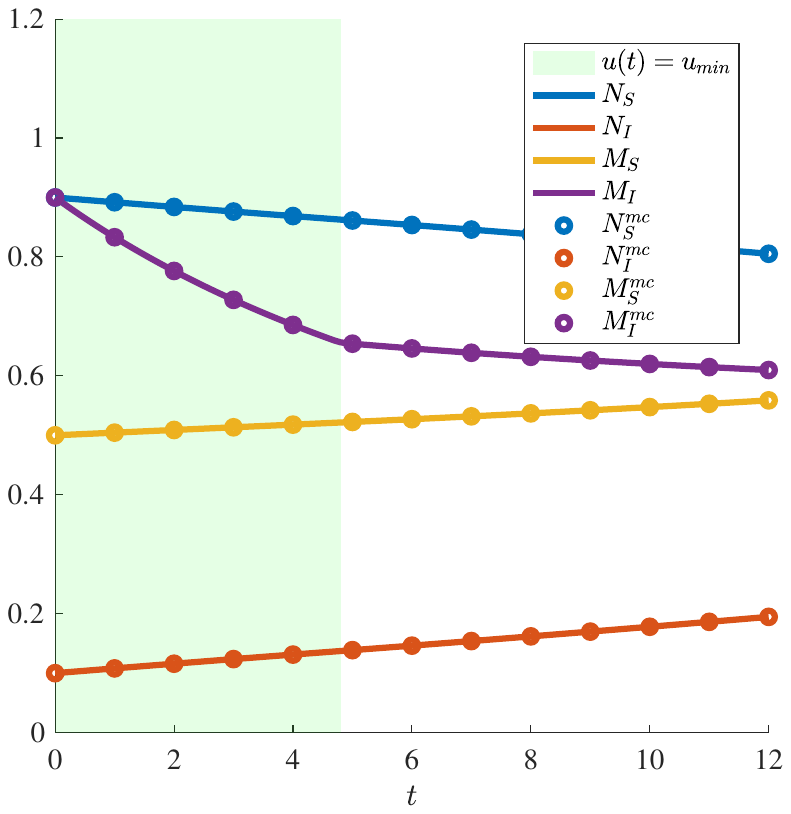}}
\subcaptionbox{\label{SIocp_q11}} 
{\includegraphics[trim={0cm 7cm 0cm 7cm}, clip, width=.45\textwidth, height=.35\textwidth]{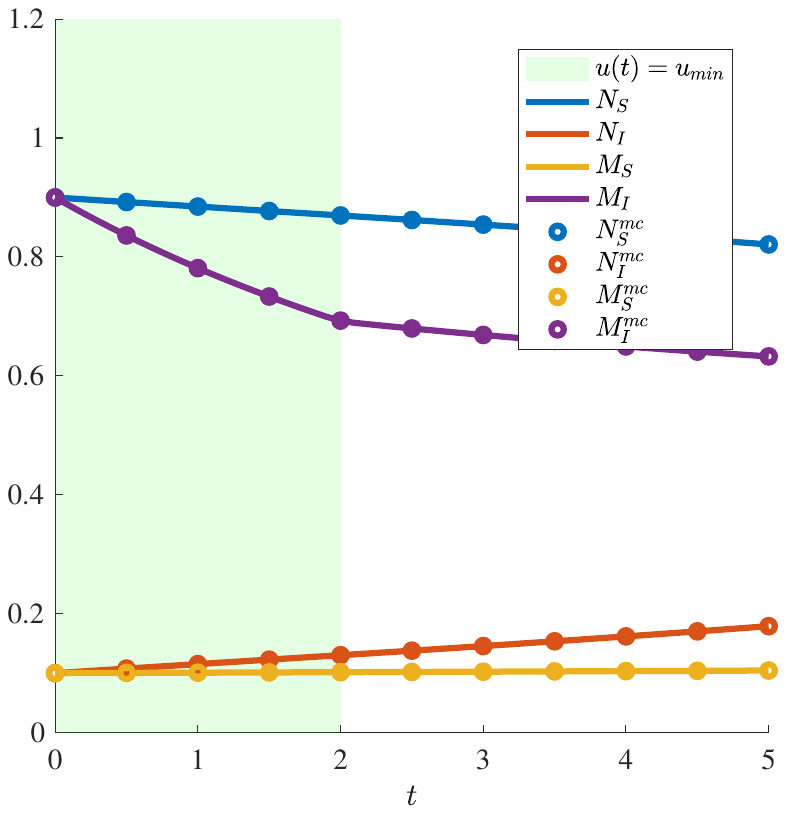}}
\hfill
\caption{{ \bf Optimal solution under assumptions of Theorems \eqref{th1} and \eqref{th2}.} Optimal trajectories of $N_S$, $N_I$, $M_S$ (the mean of resistance level in susceptible compartment, i.e $M_S=P_S/N_S$ and $M_I$ (the mean viral load in infected population, i.e. $M_I=P_I/N_I$) obtained by solving numerically the optimal control problem \eqref{ocp2} (continuous line) and the individual-based dynamics \eqref{eq:microdynlaws}-\eqref{eq:T} (circular markers) inserting the optimal control. On the left, the implementation is carried out such that the parameters involved met assumptions of Theorem \eqref{th1}, on the right such that the assumptions of Theorem \eqref{th2} are met. The green backround corresponds to the time interval where the control is \emph{active}, i.e. $u(t)=u_{min}=0 $, otherwise $u(t)=u_{max}$. }
\label{fig:SIocp_beta_q11=0}
\end{figure}
Numerical implementations show that the optimal control is characterized by \eqref{def:u_opt} even over extended  time horizon. See the Appendix for additional numerical results and all the details about numerical settings.  
\section*{Discussion and conclusion}
In this work, an optimal control problem based on an epidemic model that incorporates heterogeneity in viral load and resistance is proposed. By interpreting vaccination as an intervention targeting the average viral load acquired by individuals with the weakest immune systems, an optimal control, at least for a sufficiently small time horizon, is characterized. The analytical results show that the optimal control depends on the initial conditions and the parameters characterizing the epidemic. In certain settings, it is optimal to reduce the size of the infection by vaccinating the population at the early stages. From a phenomenological perspective, it is crucial to highlight a significant finding: the optimal control, regardless of the parameters characterizing the epidemic, does not depend on pharmaceutical availability but rather on the efficiency of the control, represented by the parameters $u_{max}$ and $u_{min}$. Furthermore, the characterization of the control, according to the assumptions of the two theorems, suggests that the intervention should be concentrated in the early stages, independent of pharmaceutical availability. These results are consistent with existing literature that investigates optimal vaccination using different approaches, as for example in  \cite{lee2011}, \cite{pinho2015}, \cite{biswas2014}, \cite{lre2026}.
\section*{Aknowledgements}
The author gratefully acknowledges Tommaso Lorenzi and Andrea Tosin for fruitful discussions during the PhD program, which have contributed to the development of this work.





\section*{Appendix}
\subsection*{Proof of Theorem \eqref{th1}}
Recalling that in case $q(1,1)=0$, $P_S \equiv P_{S,0}$ for every $t \in [0,T]$, the adjoint variable $\lambda_2$ disappears, the ODEs systems
 \beq \label{controlled:odeNiPsPix}
\begin{cases}
\displaystyle{\dot N_I=P_I \bigl[q(0,1)(1-N_I-P_S)+q(1,1)P_S\bigr]},
\\
\displaystyle{\dot P_S =-q(1,1)\,P_I\,P_S},
\\
\displaystyle{\dot P_I=P_I \bigl [ u(t)\,q(0,1)\,(1-N_I-P_S)+\overline Q(1,1)\, q(1,1)\,P_S\bigr]},
\\
\displaystyle{\dot x=u_{max}-u(t)}.
\end{cases}
\eeq
and
\beq \label{eq:adj}
\resizebox{.95\textwidth}{!}{$
\begin{cases}
\displaystyle{\dot \lambda_1= q(0,1)\,P_I\, \lambda_1 + q(0,1)\,u(t)\,P_I\,\lambda_3 },
\\
\displaystyle{\dot \lambda_2 =  [q(0,1)-q(1,1)]\,P_I\,\lambda_1 + q(1,1)\,P_I\,\lambda_2 + [q(0,1)\,u(t)-q(1,1)\,\overline Q(1,1) ]\,P_I\,\lambda_3   },
\\
\displaystyle{\dot \lambda_3 = [ -q(0,1)\,(1-N_I-P_S)-q(1,1)\,P_S] \lambda_1+q(1,1)\,P_S\,\lambda_2+-[q(0,1)\,u(t)\,(1-N_I-P_S)+q(1,1)\,\overline Q(1,1)\,P_S ]\, \lambda_3}
\\
\displaystyle{\dot \lambda_4= 0 },
\end{cases}$}
\eeq
can be rewritten as
\beq \label{eq:NiPil1l3}
\begin{dcases}
\dot N_I=q(0,1)\,P_I\,(1-N_I-P_{S,0}),
\\
\dot P_I=q(0,1)\,u(t)\,P_I\,(1-N_I-P_{S,0}),
\\
\dot \lambda_1=q(0,1)\,P_I\, \lambda_1+q(0,1)\,u(t)\,P_I \, \lambda_3,
\\
\dot \lambda_3=-q(0,1)\,(1-N_I-P_{S,0})\, \lambda_1 -q(0,1)\, u(t)\,(1-N_I-P_{S,0})\, \lambda_3
\end{dcases}
\eeq
complemented by the boundary conditions 
\beq \label{bc:NiPil1l3}
N_I(0)=N_{I,0}, \quad P_I(0)=P_{I,0}, \quad \lambda_1(T)=\lambda_0>0, \quad \lambda_3(T)=0.
\eeq
Given that $\lambda_4\equiv \beta$ for every $t$, at this stage we can omit the equation for $\lambda_4$. Since we have already noted that $\lambda_0\neq 0$, we assume the normalization $\lambda_0=1$ and we rewrite the equations \eqref{eq:NiPil1l3}$_3$-\eqref{eq:NiPil1l3}$_4$ by reversing the time dependence, i.e. $t \to -t+T$ \footnote{For convenience, at this stage, we introduce this change of variables so as to convert the original problem with a prescribed terminal condition into an equivalent Cauchy problem with prescribed initial data.
} as
\beq \label{eq:l1l3}
\begin{dcases}
\dot \lambda_1=-a(t)\,\lambda_1-a(t)\,u(t)\, \lambda_3,
\\
\dot \lambda_3=b(t)\,\lambda_1 +b(t)\,u(t)\, \lambda_3,
\end{dcases}
\eeq
where $a(t)\coloneqq q(0,1)\,P_I(t)$ and $b(t)\coloneqq q(0,1)\,(1-N_I(t)-P_{S,0})$, and complemented by the initial conditions $\lambda_1(0)=1$ and $\lambda_3(0)=0$. From now on, we omit the time-dependence and we rename in the reminder of the proof $q\coloneqq q(0,1)$. We aim to prove the statement by following these steps: 
\begin{itemize}
\item[(I)] we will provide lower and upper bounds for the functions involved in system \eqref{eq:l1l3}, i.e. $a(t)$ and $b(t)$, using equations \eqref{eq:NiPil1l3}$_{1,2}$; 
\item[(II)] we will study the evolution of the functions $\lambda_1$ and $\lambda_3$ in the right neighborhood of $t=0$, from the system \eqref{eq:l1l3}; 
\item[(III)] we will attempt to characterize the optimal control around $t=0$ using the switch function $\phi(t)$; 
\item[(IV)] finally, we will seek a condition among the parameters involved in the system \eqref{eq:NiPil1l3}, the initial conditions and the time horizon $T$ that is sufficient to ensure the characterization of the optimal control previously found.
\end{itemize}

\par
\emph{Step I. } First of all, we establish an upper and a lower bound for each functions $a$ and $b$. From equation \eqref{eq:NiPil1l3}$_2$, $P_I$ is monotone increasing, hence $P_I \ge P_{I,0} \Rightarrow a=q\,P_I \ge q\,P_{I,0}$. Dividing equation \eqref{eq:NiPil1l3}$_2$ by \eqref{eq:NiPil1l3}$_1$ and since $\dot N_I>0$,
\beq
\frac{\dot P_I}{\dot N_I}=u(t)\le u_{max} \Longrightarrow P_I-P_{I,0} \le u_{max}\,(N_I-N_{I,0}).
\eeq
By Proposition $3.2$ in \cite{tea23}, $N_I \to 1-P_{S,0}$ as $t \to \infty$, and $N_I$ is monotone increasing, then $N_I \le 1-P_{S,0}$ for every $t \ge 0$. Hence,
\beq
P_I \le P_{I,0}+u_{max}\,(1-P_{S,0}-N_{I,0}) \eqqcolon \overline P.
\eeq
Thus, we can determine both a lower and an upper bound for the function $a$, i.e.
\beq
q\,P_{I,0}\le a \le q\, \overline P.
\eeq
Given that $1-N_I-P_S=N_S(1-M_S)>0$ implies $b>0$, and due to the decreasing monotonicity of the function $N_S$, it follows that $N_S-P_{S,0}<N_{S,0}-P_{S,0}$, hence
\beq \label{boundB}
0< b \le q\, (1-N_{I,0}-P_{S,0}) \eqqcolon q\, \overline B.
\eeq
\\
\par
\emph{Step II. }Now, we study the evolution of the adjoint variables $\lambda_1$, $\lambda_3$. In particular, from \eqref{eq:l1l3}$_2$, $\dot \lambda_3(0)=b\,\lambda_1(0) >0$,
from which it follows that there exists $\e >0$ such that $\lambda_3$ is monotone increasing for every $t \in [0,\e)$. From the equations \eqref{eq:l1l3}, if $\lambda_3$ would vanish for the first time before $\lambda_1$ vanishes, i.e. if there exists $\tilde t>0$ such that $\lambda_3(\tilde t)=0$, $\lambda_3(t)>0$ for every $t \in (0,\tilde t)$ and $\lambda_1(t)>0$ for every $t\in [0,\tilde t]$, then $\dot \lambda_3(\tilde t)=b\,\lambda_1(\tilde t)>0$,
that is a contradiction. Hence, it follows that, if $\lambda_1>0$ then $\lambda_3 >0$ within the time interval. Moreover, from equation \eqref{eq:l1l3}$_1$, when $\lambda_1,\lambda_3>0$
\beq
\dot \lambda_1=-a\, \lambda_1-a\,u\, \lambda_3 <0,
\eeq 
i.e. $\lambda_1$ is strictly monotone decreasing, hence there exists $\tilde t>0$ such that $\lambda_1(\tilde t)=0$. We have proved that, for a sufficiently small $T^*$, we have $\lambda_1,\lambda_3 >0$ for every $t \in (0,T^*)$. \\
\par
\emph{Step III. }Recalling the definition of the switch function 
\beq \label{def:switch}
\phi(t)=\lambda_3\,P_I\,q(0,1)\,(1-N_I-P_S)-\lambda_4.
\eeq
we compute its derivative and we evaluate its sign by restoring the original time direction of the functions $\lambda_1$ and $\lambda_3$.After straightforward computations, it follows that
\begin{align*}
\dot \phi(t)=&(\dot \lambda_3 ) \, q\, P_I \,(1-N_I-P_{S,0})+\lambda_3\, q\, (\dot P_I)\,(1-N_I-P_{S,0})+\lambda_3\,q\,P_I\,(-\dot N_I)\\
=&\Bigr[ -q\,(1-N_I-P_{S,0})\,\lambda_1-q\,u(t)\,(1-N_I-P_{S,0})\,\lambda_3 \Bigl]\, q\,P_I\,(1-N_I-P_{S,0})+\\
&\lambda_3\,q\,P_I\, u(t)\,q\,(1-N_I-P_{S,0})^2+\lambda_3\, q\,P_I\,\Bigr[-P_I\,q\,(1-N_I-P_{S,0})\Bigl]\\
=&(1-N_I-P_{S,0})\,q\,P_I\,\Bigr[-q\,(1-N_I-P_{S,0})\,\lambda_1-q\,P_I\,\lambda_3 \Bigl]<0,
\end{align*}
since $\lambda_1,\lambda_3 > 0$ for every $t \in (0,T^*)$. 
It follows that the switch function is monotone decreasing, at least in $[0,T^*]$, when $\lambda_1>0$, $\lambda_3\ge 0$. Considering that it must vanish at least once, otherwise the control is not admissible by the $L^1$ constraint, i.e. 
\beq\label{ass:L1}
C= \int_0^{T} [u_{max}-u(t)] \di t,
\eeq
the sign switches from positive to negative, i.e.
\beq
\begin{dcases}
\phi(t)>0 \quad &\text{ for } t \in [0, \tilde T),
\\
\phi(t)<0 \quad &\text{ for } t \in (\tilde T, T^*],
\end{dcases}
\eeq
and, as a consequence, 
\beq \label{u:switch}
u(t)=
\begin{dcases}
u_{min} \quad &\text{ for } t \in [0, \tilde T) ,
\\
u_{max} \quad &\text{ for } t \in (\tilde T,T^*].
\end{dcases}
\eeq
where $\displaystyle \tilde T=\frac{C}{u_{max}-u_{min}}$. \\
\par
\emph{Step IV. }Finally, the goal is to establish a sufficient condition between the parameters involved in the system \eqref{eq:l1l3} and $T^*$ in order to guarantee that $\lambda_1,\lambda_3 >0$ for every $t\in (0,T^*)$. We have already argued in Step 2 that $\lambda_1>0$ for every $t\in (0,T^*)$ is sufficient to guarantee that $\lambda_3>0$ within the same time interval. We consider again the ODEs system 
\eqref{eq:l1l3} where the time dependence is reversed.
Since $\lambda_1,\lambda_3>0$ lead to $\dot \lambda_3 >0$ for every $t \in (0,T^*)$, it follows that
\beq
\max_{t \in[0,T^*]} \lambda_3(t)=\lambda_3(T^*) \eqqcolon \overline \lambda_3.
\eeq
For notational convenience, from now on we identify $T^*$ with $T$. From equation \eqref{eq:l1l3}$_1$, for every $t \in [0,T]$,
\begin{equation*}
\dot \lambda_1=-a\,\lambda_1-a\, u\, \lambda_3 \ge -q \, \overline P \lambda_1-q\, \overline P\, u_{max} \overline \lambda_3 
\Longrightarrow \lambda_1(t) \ge e^{-q \overline P t} \Biggl [ \lambda_1(0)-q\, \overline P \, u_{max} \, \overline \lambda_3 \int_0^t e^{q \overline P  \tau} \di \tau \Biggr ],
\end{equation*}
thus
\beq \label{bound:l1}
\lambda_1(T) \ge e^{-q \overline P T} \Bigl [ 1- u_{max} \, \overline \lambda_3\,  (e^{q \overline P T}-1) \Bigr ].
\eeq
From equation \eqref{eq:l1l3}$_2$, we could establish an upper bound for $\overline \lambda_3$. In particular, since $\lambda_1, \lambda_3 >0$ and $\lambda_1$ is strictly monotone decreasing, then $\lambda_1(t) <\lambda_1(0)=1$, and from the bounds on the function $b$, i.e \eqref{boundB}, yields
\begin{equation*}
\dot \lambda_3=b\, \lambda_1 +b\, u\,\lambda_3 \le q\, \overline B+q\,\overline B\,u_{max}\,\lambda_3 
\Longrightarrow \lambda_3(t) \le e^{q \overline B u_{max} t} \frac{1}{u_{max}} \Bigr (1-e^{-q\overline B u_{max} t} \Bigl),
\end{equation*}
hence
\beq \label{bound:maxl3}
\overline \lambda_3 \le \frac{1}{u_{max}} \Bigr( e^{q\, \overline B \, u_{max}\, T}-1 \Bigl).
\eeq
Inserting \eqref{bound:maxl3} in \eqref{bound:l1}, we obtain $\displaystyle \lambda_1(T) \ge e^{-q\, \overline P\, T} \Bigr[1-\Bigr( e^{q\, \overline B \, u_{max}\, T}-1 \Bigl)\Bigr(e^{q\, \overline P \, T}-1 \Bigl) \Bigl]$. 
We remark that $\lambda_3(T)$ is still strictly positive even if $\lambda_1(T)=0$ since $\dot\lambda_3=b\,\lambda_1+b\,u\,\lambda_3>0$
as long as $\lambda_1>0$ so that $\lambda_3$ remains strictly positive by continuity up to $t=T$.
And this is enough to guarantee that $\dot \phi(t)<0$ for every $t\in [0,T]$. Imposing that the lower bound for $\lambda_1(T)\ge 0$ the statement of the theorem is obtained. 
\subsection*{Proof of Theorem \eqref{th2}}
We start by rewriting the ODEs system \eqref{eq:adj} complemented by the transversality conditions at the final time, i.e.
\beq\label{adja}
\lambda_1(T)=1 \ge 0,
\quad \lambda_2(T)=\lambda_3(T)=0,
\eeq
reversing the time-dependence, thus
\beq \label{eq:l1l2l3}
\begin{dcases}
\dot \lambda_1=-a\,\lambda_1-a\,u\,\lambda_3,\\
\dot \lambda_2=-d\,\lambda_1-e\,\lambda_2-f\, \lambda_3,\\
\dot \lambda_3=g\,\lambda_1-h\,\lambda_2 +j\,\lambda_3,\\
\end{dcases}
\eeq
with 
\beq \label{adj:bc}
\lambda_1(0)=1, \quad \lambda_2(0)=\lambda_3(0)=0,
\eeq 
where the functions $a$, $d$, $e$, $f$, $g$, $h$ and $j$ are defined as
\beq \label{def:ADEFGHJ} 
a \coloneqq q(0,1)\,P_I,\quad  d \coloneqq \Bigr[q(0,1)-q(1,1)\Bigl]\,P_I, \quad e \coloneqq q(1,1)\,P_I, \quad f \coloneqq \Bigr[ q(0,1)\,u-q(1,1)\,\overline Q(1,1) \Bigl]\,P_I,
\eeq
\beq \label{def:ADEFGHJ2}
g \coloneqq q(0,1)\,(1-N_I-P_S)+q(1,1)\,P_S,\quad h \coloneqq q(1,1)\,P_S, \quad j \coloneqq q(0,1)\,u\,(1-N_I-P_S)+q(1,1)\, \overline Q(1,1) \,P_S,
\eeq
and where the time-dependence of all functions is omitted. We will prove the statement following the same approach built in the previous Theorem.\\ \par
\emph{Step I. }Considering the increasing monotonicity of $N_I$ and $P_I$, and the decreasing monotonicity of $P_S$, see \eqref{controlled:odeNiPsPix}$_2$, and the assumption $0< q(1,1)<q(0,1) \le 1$, the latter functions lead to the following bounds
\begin{equation} \label{bound:ADEGHJ} 
0<q(0,1)\, P_{I,0} \le  a \le q(0,1)\, \overline P, \quad 0<\Bigr [q(0,1)-q(1,1)\Bigl]\, P_{I,0} \le  d \le  [q(0,1)-q(1,1)]\, \overline P, 
\end{equation}
\beq \label{bound:ADEGHJ2}
 0<q(1,1)\, P_{I,0} \le  e \le q(1,1)\, \overline P,  \quad  0<g \le q(0,1)\,(1-N_{I,0})+q(1,1)\,P_{S,0} \eqqcolon \overline G,
\eeq
\beq \label{bound:ADEGHJ3}
0 < h \le q(1,1)\,P_{S,0}, \quad 0 < j \le q(0,1)\,u_{max}\,(1-N_{I,0})+q(1,1)\,\overline Q(1,1)\, P_{S,0} \eqqcolon \overline J,
\eeq
where $\overline P=P_{I,0}+u_{max}\,(1-N_{I,0}-P_{S,0})+\overline Q(1,1)\,P_{S,0}$ (see Prop. 3.2 \cite{tea23}). 
Furthermore, we estimate the upper and lower bounds for the function $f$ defined in \eqref{def:ADEFGHJ}, i.e.
\beq
\Bigr [q(0,1)\,u_{min}-q(1,1)\,\overline Q(1,1) \Bigl]\,P_I \le f \le \Bigr [q(0,1)\,u_{max}-q(1,1)\,\overline Q(1,1) \Bigl]\,P_I.
\eeq
Since $f$ can switch the sign, i.e. $q(0,1)\,u_{min}-q(1,1)\,\overline Q(1,1)\le0$ and $q(0,1)\,u_{max}-q(1,1)\,\overline Q(1,1) \ge0$ \footnote{It is natural to assume that $u_{max}=\overline Q(0,1) \ge \overline Q(1,1)$ meaning that, $u(t)=u_{max}$ corresponds to the scenario where the control is not active, and its value is the parameter $\overline Q(0,1)$ in absence of any control. On the other side, it is natural to assume that $u_{min} \approx 0$, indeed $q(0,1)u_{min}\le q(1,1)\overline Q(1,1)$ such that the pharmaceutical intervention is more efficient to slow down the infection curve, see the analytical results in \cite{tea23}. Otherwise, the proof remains valid after suitably adapting the bounds on $f$.}, 
\beq
\Bigr [q(0,1)\,u_{min}-q(1,1)\,\overline Q(1,1) \Bigl]\,\overline P \le f \le \Bigr [q(0,1)\,u_{max}-q(1,1)\,\overline Q(1,1) \Bigl]\,\overline P,
\eeq
thus
\beq
-\tilde F \coloneqq- \Bigr [q(0,1)\,u_{max}-q(1,1)\,\overline Q(1,1) \Bigl]\,\overline P \le -f \le - \Bigr [q(0,1)\,u_{min}-q(1,1)\,\overline Q(1,1) \Bigl]\,\overline P \eqqcolon \overline F, \label{bound:F}
\eeq
with $\tilde F$, $\overline F > 0$. \\ \par
\emph{Step II. } We observe that 
\beq
\dot \lambda_1 (0)=-a <0, \quad \quad \dot \lambda_2(0)=-d<0, \quad \quad \dot \lambda_3(0)=g >0,
\eeq
from which it follows that $\lambda_1$ and $\lambda_2$ are strictly decreasing and $\lambda_3$ is strictly increasing in a neighborhood of $t=0$. Due to the initial conditions \eqref{adj:bc}, there exists $T^*>0$ such that $\lambda_1(t), \lambda_3(t) >0$ and $\lambda_2(t)<0$ for every $t \in (0,T^*)$. \\ \par
\emph{Step III. } We compute the first derivative of the switch function \eqref{def:switch} and evaluate its sign by restoring the original time direction of the functions $\lambda_1$, $\lambda_2$ and $\lambda_3$. We obtain
\beq 
\dot \phi(t)=\dot \lambda_3 \, q(0,1) \, P_I \,(1-N_I - P_S)+\lambda_3 \, q(0,1)\, \dot P_I \,(1-N_I-P_S)+\lambda_3 \, q(0,1) \, P_I\, (-\dot N_I-\dot P_S ),
\eeq 
which, along with \eqref{controlled:odeNiPsPix} and \eqref{eq:adj}, leads to
\beq \label{def:mon_switch}
\resizebox{.93\textwidth}{!}{$ \displaystyle
\dot \phi(t)=\Bigr \{-\Bigr [q(0,1)\,(1-N_I-P_S)+q(1,1)\,P_S \Bigl]\,\lambda_1+q(1,1)\,P_S\,\lambda_2 \Bigl \}\,q(0,1)\,P_I\,(1-N_I-P_S)-q(0,1)^2\,(1-N_I-P_S)\,P_I^2\,\lambda_3 < 0$},
\eeq
for every $t \in [0,T^*]$, since $\lambda_1(t), \lambda_3(t) >0$ and $\lambda_2(t)<0$ for every $t \in (0,T^*)$ and $\lambda_1(T^*)>0$, $\lambda_2(T^*),\lambda_3(T^*)=0$. Moreover, it is trivial to check that, for a sufficiently small $T^*$, $(\lambda_1(0),\lambda_2(0),\lambda_3(0)) \neq 0$. By arguments already discussed in the third step of the latter Theorem, we can conclude that \eqref{u:switch} holds true. \\
\par

\emph{Step IV. }Now, we aim to establish a condition between the parameters involved in the system \eqref{eq:l1l2l3} and $T^*$ such that it is sufficient to guarantee that $\lambda_1, \lambda_3 > 0$ and $\lambda_2 < 0$ for every $t\in (0,T^*)$. 
By absurd, we assume that the first function that will vanish is $\lambda_3$, i.e. there exists $\tilde t>0$ such that $\lambda_3(\tilde t) =0$, $\lambda_3(t)>0$ for every $t \in (0,\tilde t)$, and $\lambda_1(t) >0$, $\lambda_2(t)<0$ for every $t \in (0, \tilde t)$. From this assumption and from equation \eqref{eq:l1l2l3}$_3$, we get
\beq
\dot \lambda_3(\tilde t)= g\,\lambda_1(\tilde t)-h\, \lambda_2 (\tilde t) >0, 
\eeq
that is an absurd. Among $\lambda_1$, $\lambda_2$ and $\lambda_3$, it folows that $\lambda_3$ can not be the first to vanish. 
For notational convenience, from now on we rename $T^*=T$. 
Now, we assume that $\lambda_1$ vanishes before $\lambda_2$ and our aim is to find a condition involving $T$, the initial conditions $N_{I,0}$, $P_{S,0}$, $P_{I,0}$ and the parameters $q(1,1)$, $q(0,1)$ and $\overline Q(1,1)$ such that $\lambda_1(t)> 0$ for every $t \in (0,T)$. 
Then, in similar way, we assume that $\lambda_2$ would vanish before $\lambda_1$ and we will establish a condition ensuring the non-positivity of $\lambda_2$. If the two conditions hold for every $t \in (0,T)$, then $\lambda_1, \lambda_3 > 0$ and $\lambda_2 < 0$ in $(0,T)$.\\
By equation \eqref{eq:l1l2l3}$_1$, when $\lambda_1>0$ and $\lambda_3>0$, $\lambda_1$ is strictly decreasing, hence $\lambda_1(t)<\lambda_1(0)=1$ for every $t \in (0,T]$. Then, by \eqref{eq:l1l2l3}$_3$ and the estimation \eqref{bound:ADEGHJ}-\eqref{bound:ADEGHJ2}, we get
\beq
\dot \lambda_3 =g\,\lambda_1-h\, \lambda_2 + j\,\lambda_3 \le \overline G+ q(1,1)\,P_{S,0}\, \overline \lambda_2 + \overline J \, \lambda_3,
\eeq
where 
\beq
\overline \lambda_2 \coloneqq \max_{t \in [0,T]} \left | \lambda_2(t) \right |.
\eeq
The latter inequality, along with \eqref{adj:bc}, leads to 
$\displaystyle \lambda_3(t) \le \frac{\overline G+q(1,1)\, P_{S,0}\, \overline \lambda_2}{\overline J} \Bigr(e^{\overline J t}-1 \Bigl).$
Hence, since by equation \eqref{eq:l1l2l3}$_3$, $\lambda_3$ is monotone increasing, yields
\beq \label{bound:l3}
\overline \lambda_ 3\coloneqq \max_{t\ \in [0,T]} \lambda_3(t)=\lambda_3(T)  \le \frac{\overline G+q(1,1)\, P_{S,0}\, \overline \lambda_2}{\overline J} \Bigr(e^{\overline J T}-1 \Bigl). 
\eeq
Now, from equation \eqref{eq:l1l2l3}$_2$ along with 
\eqref{bound:ADEGHJ} and \eqref{bound:F}, we obtain
\beq
\dot \lambda_2=-d\,\lambda_1-e\, \lambda_2-f\,\lambda_3 \ge -d-e\,\lambda_2-f\overline \lambda_3 \ge -[q(0,1)-q(1,1)]\, \overline P-q(1,1)\, \overline P\, \lambda_2 -\tilde F \,\overline \lambda_3,
\eeq
from which it follows that
\beq  \label{bound:l2}
 \lambda_2(t) \ge \frac{\Bigr[q(0,1)-q(1,1) \Bigl] \overline P+\tilde F \, \overline \lambda_3}{q(1,1)\,\overline P} \Bigr(e^{-q(1,1)\overline P t}-1 \Bigl) \quad \Longrightarrow \quad 
\overline \lambda_2 \le \frac{\Bigr[q(0,1)-q(1,1) \Bigl ] \overline P+\tilde F \, \overline \lambda_3}{q(1,1)\,\overline P} \Bigr(1-e^{-q(1,1)\overline P T} \Bigl).
\eeq
The latter inequality along with \eqref{bound:l3}, leads to 
\begin{equation*}
\overline \lambda_3 \le  \frac{1}{\overline J} \Biggr \{ \overline G+q(1,1)\,P_{S,0} \frac{  [q(0,1)-q(1,1) ]\, \overline P+\tilde F\, \overline \lambda_3}{q(1,1)\, \overline P} \Bigr(1-e^{-q(1,1)\overline P T}\Bigl) \Biggl \}\Bigr(e^{\overline J T}-1 \Bigl )
\end{equation*}
\beq\label{cond:maxl3}
\Longrightarrow  \overline \lambda_3 \Biggr \{1- \frac{P_{S,0}}{\overline J \, \overline P} \tilde F \Bigr(1-e^{-q(1,1) \overline P T} \Bigl ) \Bigr (e^{\overline J T}-1 \Bigl) \Biggl \} \le 
\frac{1}{\overline J} \Biggr \{ \overline G+ P_{S,0} \Bigr [q(0,1)-q(1,1) \Bigl ] \Bigr(1-e^{-q(1,1)\overline P T}\Bigl) \Biggl \} \Bigr( e^{\overline J T}-1 \Bigl).
\eeq
Hence, if the condition 
\beq \label{sign:l3}
\overline J- \frac{P_{S,0}}{\overline P} \tilde F \Bigr(1-e^{-q(1,1) \overline P T} \Bigl ) \Bigr (e^{\overline J T}-1 \Bigl)>0, 
\eeq
is satisfied, then
\beq \label{max:l3}
\overline \lambda_3 \le \frac{  \overline G+ P_{S,0} \Bigr [q(0,1)-q(1,1) \Bigl] \Bigr(1-e^{-q(1,1)\overline P T}\Bigl)}{\overline J- \frac{P_{S,0}}{\overline P} \tilde F \Bigr(1-e^{-q(1,1) \overline P T} \Bigl ) \Bigr (e^{\overline J T}-1 \Bigl)}   \Bigr( e^{\overline J T}-1 \Bigl) \eqqcolon \lambda_3^*,
\eeq
otherwise \eqref{cond:maxl3} is always satisfied and the estimation of the upper bound of $\overline \lambda_3$ falls. 
The equation \eqref{eq:l1l2l3}$_1$ along with the bounds estimated in \eqref{bound:ADEGHJ}, leads to
\beq 
\dot \lambda_1 =-a\,\lambda_1-a\,u\,\lambda_3 \ge-q(0,1)\, \overline P \, \lambda_1-q(0,1)\, \overline P\, u_{max} \, \overline \lambda_3  \quad \Longrightarrow \quad  \lambda_1(T) \ge e^{-q(0,1)\overline P T} \Bigr[ 1-u_{max} \, \overline \lambda_3 \Bigr( e^{q(0,1) \overline P T}-1\Bigl) \Bigl]
\eeq
thus, along with  \eqref{max:l3}, 
\beq  \label{min:l1}
\lambda_1(T) \ge e^{-q(0,1)\overline P T} \Biggr[ 1-u_{max} \, \overline \lambda_3 \Bigr( e^{q(0,1) \overline P T}-1\Bigl) \Biggl ] \ge e^{-q(0,1)\overline P T} \Biggr[ 1-u_{max} \, \lambda_3^* \Bigr( e^{q(0,1) \overline P T}-1\Bigl) \Biggl ].
\eeq
from wich it follows that
\beq \label{cond:l1}
  u_{max} \,  \lambda_3^*                 \Bigr( e^{q(0,1) \overline P T}-1\Bigl)            \le 1 \quad \Longrightarrow \quad \lambda_1(t) \ge 0 \quad \text{ for every } t \in [0,T].
\eeq
Indeed, we get a condition that ensures the non-negativity of $\lambda_1$ in case when $\lambda_1$ would vanish before $\lambda_2$ and $\lambda_3$, in the interval $(0,T)$.  
In similar way, we want to establish a condition that ensures that $\lambda_2 \le 0$ in the case where $\lambda_2$ would vanish before $\lambda_1$. From \eqref{eq:l1l2l3}$_2$, \eqref{bound:ADEGHJ} and \eqref{bound:F} and defining 
\beq
 \tilde \lambda_1 \coloneqq \min_{t \in [0,T]} \lambda_1(t),
\eeq
we get 
\begin{equation*}
\dot \lambda_2 \le - \Bigl [q(0,1)-q(1,1) \Bigr ]\,P_{I,0}\,\lambda_1-q(1,1)\,P_{I,0}\, \lambda_2 + \overline F\,\lambda_3 \le - \Bigl [q(0,1)-q(1,1) \Bigr ]\,P_{I,0}\,\tilde \lambda_1-q(1,1)\,P_{I,0}\, \lambda_2 + \overline F\,\overline \lambda_3,
\end{equation*}
hence, along with \eqref{adj:bc}, yields
\begin{equation*}
\lambda_2(T) 
\le e^{-q(1,1)P_{I,0}T}  \Biggl \{ -\Bigl [q(0,1)-q(1,1) \Bigr ]\,P_{I,0}\,\tilde \lambda_1 +\overline F\, \overline \lambda_3 \Biggr \} \frac{1}{q(1,1)\,P_{I,0}}\Bigl (e^{q(1,1)P_{I,0}T}-1 \Bigr ).
\end{equation*}
The condition
\beq \label{nonpos:l2}
-\Bigl [q(0,1)-q(1,1) \Bigr] \, P_{I,0}\tilde \lambda_1+\overline F \,\overline \lambda_3 \le 0
\eeq
ensures that $\lambda_2(t) \le 0$ for every $t \in [0,T]$. 
Recalling the inequality \eqref{min:l1} and the fact that $\lambda_1$ is monotone decreasing when $\lambda_1, \lambda_3 >0$, we get 
\beq
\label{min:l1_new}
\tilde \lambda_1 \ge e^{-q(0,1)\overline P T} \Biggr[ 1-u_{max} \, \overline \lambda_3 \Bigr( e^{q(0,1) \overline P T}-1\Bigl) \Biggl ],
\eeq
that, inserted in \eqref{nonpos:l2} leads to
\begin{equation}
\begin{split}
&-\Bigl [q(0,1)-q(1,1) \Bigr] \, P_{I,0}\tilde \lambda_1+\overline F \,\overline \lambda_3 \le-\Bigl[q(0,1)-q(1,1)\Bigr]\,P_{I,0}\,e^{-q(0,1)\overline P T} \Bigl [1-u_{max}\,\overline \lambda_3 \Bigl( e^{q(0,1)\overline P T}-1 \Bigr) \Bigr ]+\overline F\,\overline \lambda_3.   \\
\end{split}
\end{equation}
Imposing the right-hand side of the latter inequality non-positive, i.e.
\beq \label{cond:l3bar}
\overline \lambda_3 \Biggl \{ \overline F+\Bigl[q(0,1)-q(1,1) \Bigr]\,P_{I,0}\,e^{-q(0,1)\overline P T}\,u_{max}\Bigl(e^{q(0,1)\overline P T}-1\Bigr) \Biggr \} - \Bigl[q(0,1)-q(1,1)\Bigr]\,P_{I,0}\, e^{-q(0,1)\overline P T}         \le 0,
\eeq
 is sufficient to guarantee that \eqref{nonpos:l2} holds true.
It implies that, if
\beq \label{cond:l2_new}
\lambda_3^* \le \frac{\Bigl[q(0,1)-q(1,1)\Bigr]\,P_{I,0}\, e^{-q(0,1)\overline P T}}{\overline F+\Bigl[q(0,1)-q(1,1) \Bigr]\,P_{I,0}\,e^{-q(0,1)\overline P T}\,u_{max}\Bigl(e^{q(0,1)\overline P T}-1\Bigr)},
\eeq
where $\lambda_3^*$ is defined via \eqref{max:l3}, then the condition \eqref{cond:l3bar} holds true, hence also \eqref{nonpos:l2} does and $\lambda_2 \le 0$ for every $t \in [0,T]$. Since we can not establish a priori which functions between $\lambda_1$ and $\lambda_2$ can vanish before, we compare the two conditions \eqref{cond:l1} and \eqref{cond:l2_new}, one that guarantees the positive sign of $\lambda_1$ and the other the negative sign of $\lambda_2$ and we choose the more restrictive condition that ensures $\lambda_1>0$ and $\lambda_2<0$ for every $t\in (0,T)$. We recall the fact that, in guaranteeing $\lambda_1>0$ and $\lambda_2<0$, automatically yields $\lambda_3>0$ in $(0,T)$. 
Renaming the quantity
\beq
k \coloneqq u_{max}\,\Bigl(e^{q(0,1)\overline P T}-1 \Bigr),
\eeq
we notice that the condition \eqref{cond:l1} leads to 
\beq \label{cond:fin1}
\lambda_3^* \le \frac{1}{u_{max}\,\Bigl(e^{q(0,1)\overline P T}-1 \Bigr)}=\frac{1}{k} \quad \Longrightarrow \quad \frac{1}{\lambda_3^*} \ge k,
\eeq
and from \eqref{cond:l2_new} we have
\begin{equation}\label{cond:fin2}
\frac{1}{\lambda_3^*} \ge \frac{\overline F+\Bigl[q(0,1)-q(1,1) \Bigr]\, P_{I,0}\,e^{-q(0,1)\overline P T}\,k}{\Bigr[q(0,1)-q(1,1) \Bigl]\,P_{I,0}\,e^{-q(0,1)\overline P T}}
=\frac{\overline F}{\Bigr[q(0,1)-q(1,1) \Bigl]\,P_{I,0}\,e^{-q(0,1)\overline P T}}+k \ge k.\\
\end{equation}
We obtain that, if the condition \eqref{cond:fin2} is satisfied, also \eqref{cond:fin1} does. Indeed, the condition \eqref{cond:l2_new} ensures that \eqref{cond:l1} holds true. 
We remark that the upper bound \eqref{max:l3} for $\overline\lambda_3$ holds independently of whether $\lambda_1$ or $\lambda_2$ vanishes first, since its derivation only relies on \eqref{bound:l3}-\eqref{bound:l2}, valid under conditions \eqref{sign:l3}.

Finally, reinserting the expression for $\lambda_3^*$ given by \eqref{max:l3}, in \eqref{cond:l2_new}, we obtain the condition 
\begin{equation*}
\frac{\overline G+P_{S,0}\,\Bigl [q(0,1)-q(1,1) \Bigr] \Bigl(1-e^{-q(1,1)\overline P T} \Bigr )}{\overline J-\frac{P_{S,0}}{\overline P}\tilde F \Bigl(1-e^{-q(1,1) \overline P T}\Bigr)\Bigl(e^{\overline JT}-1 \Bigr)}\Bigl(e^{\overline J T}-1 \Bigr) \le 
\frac{\Bigr [q(0,1)-q(1,1) \Bigl]\,P_{I,0}\,e^{-q(0,1)\overline PT}}{\overline F+\Bigr[q(0,1)-q(1,1) \Bigl]\,P_{I,0}\,u_{max}\,\Bigl(1-e^{-q(0,1)\overline P T} \Bigr)}
\end{equation*}
which along with \eqref{sign:l3}, implies that
\beq
\lambda_1(t),\lambda_3(t) > 0 \text{ and } \lambda_2 (t) < 0 \text { for every } t \in (0,T).
\eeq
Combining this with \eqref{def:mon_switch}, $\dot\phi(t)<0$ for every $t\in(0,T)$
hence $\phi$ is strictly monotone decreasing on $[0,T]$, which concludes the proof.

\subsection*{Numerical setting} \label{sec4}
The optimal control problem is solved by a discretization method. The control problem provides a large-scale nonlinear optimization problem  which is formulated within the Mathematical Programming Language (AMPL) developed by \cite{fourer1993}. We utilize the primal-dual interior point algorithm IPOPT, see \cite{wachter2006}, written in AMPL  and directly submitted in NEOS server, \cite{neos}. As integration method we use the trapezoidal rule. We choose to discretize the time horizon  $T $ with $N = 10\cdot T$ steps. We prescribe the tolerance tol $ = 10^{-8}$ in all computations.
We set $u_{max}=0.5$ and $ u_{min}=0$. 
 We fix the value $C$ in the integral constraint such that $\tilde T=2/5\cdot T$, hence $C=\tilde T\,(u_{max}-u_{min})= \tilde T/2$. This translates into the idea that the pharmaceutical source is able to bring the parameter $\overline Q(0,1)$ to $0$ for the $40\%$ of the time length.  To provide a more meaningful interpretation of the result, we present below the trajectories of the quantities $N_S$, $N_I$, $M_S$, and $M_I$, where the latter two represent the average resistance among susceptibles and the average viral load among infected individuals, respectively. These are obtained computing $M_S = P_S / N_S$ and $M_I = P_I / N_I$. In all computations, we prescribe $N_{I,0}=0.1$, $M_{I,0}=0.9$ and $M_{S,0}=0.5$ or $M_{S,0}=0.1$.\\ 
To highlight the effect of the control, the \emph{macroscopic} dynamics in the absence of any control are also implemented, specifically using the built-in ODE45 function in Matlab. \\
Controlling the quantity $\overline Q(0,1)$ translates to the idea, on a \emph{microscopic} level, of managing the interaction between a susceptible individual with resistance $r=0$ and an infected individual with viral load $v=1$. However, this type of control affects all interactions involving a highly vulnerable susceptible. Implicitly assuming that the higher the viral load of the infected, the greater the viral load acquired by the susceptible post-interaction, vaccinating the susceptible with $r=0$ such that they would acquire a viral load $u_{min}$ if they encounter a super-infected individual, means that even if they get infected by an individual with a lower viral load, they would still acquire a viral load $u_{min}$ upon infection. Therefore, the formulation of the problem translates to the idea that by controlling only the quantity $\overline Q(0,1)$, there are practical implications for vaccinating susceptibles with the weakest immune systems, ensuring they acquire a viral load $u_{min}$ following infection from any infected individual.\\
We implement the \emph{microscopic} dynamics 
through a Monte Carlo method, considering the distribution function $Q(t,v|r,v^*)$ such that its mean at each time is defined according to the control, i.e. $u(t)=\overline Q(t,0,1)$, obtained by the implementation of the optimal control problem. In particular, in all numerical simulations we prescribe $Q(t,v|r,v^*)$ as
\beq \label{def:Qbeta8}
Q(t,v|r,v^*)=
\begin{cases} 
Q_1(v|r,v^*) &\text{ if } t\in[0,T]: u(t)=u_{max} ,\\
Q_2(v|r,v^*)  &\text{ if }t\in[0,T]: u(t)=u_{min} ,\\
\end{cases}
\eeq
where $Q_1(v|r,v^*)$ is a Beta distribution function with mean 
\beq \label{def:meanQ}
\overline Q(r,v^*)= (0.05+0.45\,(1-r))\,v^*, 
\eeq
and variance $\sigma^2=0.001$, such that
\beq
\overline Q(1,1)=0.05, \quad \overline Q(0,1)=u_{max}=0.5,
\eeq
and 
\beq
Q_2(v|r,v^*)=
\begin{cases} 
\delta_0(v) &\text{ if } (r,v^*)=(0,1)  ,\\
Q_1(v|r,v^*)  &\text{ otherwise},\\
\end{cases}
\eeq
indeed $\overline Q(1,1)=0.05$ and $\overline Q(0,1)=u_{min}=0$. Moreover, the function $q$ is defined as
\beq \label{q_1}
q(r,v^*)=0.2\,(1-r)\,v^* \quad \text{ s.t. } q(1,1)=0 \text{  and  } q(0,1)=0.2,
\eeq
or as
\beq \label{q_2}
q(r,v^*)=(0.2-0.1\,r)\,v^* \quad \text{ s.t. } q(1,1)=0.1 \text{  and  } q(0,1)=0.2,
\eeq
for implementing numerical solutions such that the assumptions of Theorem 1 and Theorem 2, respectively, are met. At the microscopic level, the probability distributions $K_S(r'|r)$ and $K_I(v'|v)$ are defined as binomial distributions
\beq
K_S(r'| r)=(1-r)\,\delta_0(r')+r\,\delta_1 (r'), \quad K_I(v'| v)=(1-v)\,\delta_0(v')+v\,\delta_1(v').
\eeq
In all figures reported below, the continuous line highlights the numerical results of the \emph{macroscopic} dynamics and the circular markers the corresponding quantities obtained through Monte Carlo simulations of the \emph{microscopic} model. The green backround indicates the time interval where the optimal control is \emph{active}, i.e. $u(t)=u_{min}$, and the white one where the control is \emph{not active}, i.e. $u(t)=u_{max}$.\\
In the examples displayed below, it can be observed that, as a natural consequence of controlling the mean viral load, the evolution of the infected population $M_I$ is significantly slowed down compared with the uncontrolled case. The impact of the control on the $M_I$ curve is clearly noticeable, leading to a substantial reduction in the epidemic growth rate. We also emphasize that, due to the intrinsic nature of the SI model, which does not account for recovery or death, complete eradication of the epidemic cannot be achieved. In fact, extinction would only occur for parameter choices that entirely suppress the infection dynamics.

 \begin{figure}[H]
\centering
\subcaptionbox{\label{SIocpV9R5_beta_q11=0}}
{\includegraphics[trim={0cm 7cm 0cm 7cm}, clip, width=.45\textwidth, height=.35\textwidth]{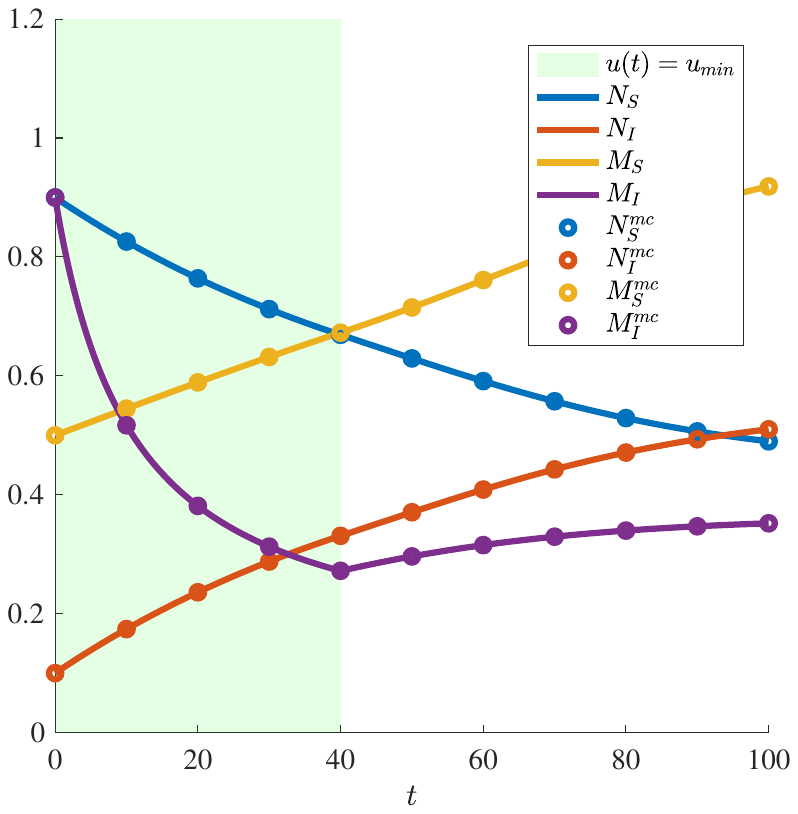}}
\subcaptionbox{\label{SIV9R9_beta_q11=0}} 
{\includegraphics[trim={0cm 7cm 0cm 7cm}, clip, width=.45\textwidth, height=.35\textwidth]{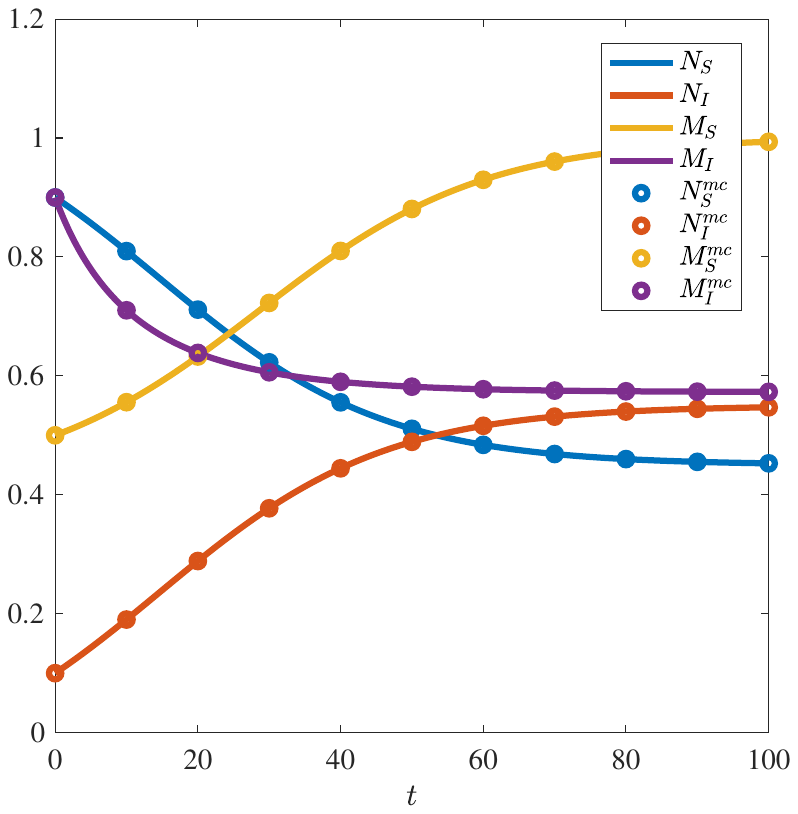}}
\hfill
\caption{{ \bf Optimal solution extending Theorem 1 and the uncontrolled dynamics.} On the right the trajectories of the functions $N_S$, $N_I$, $M_S$, and $M_I$ without any control, obtained by prescribing $q$ defined via \eqref{q_1}. On the left, the optimal trajectories obtained by implementing the corresponding optimal control problem, extending the time horizon.}
\end{figure}
 \begin{figure}[H]
\centering
\subcaptionbox{\label{SIocpV5R5_beta_q11=0}}
{\includegraphics[trim={0cm 7cm 0 7cm},clip, width=.45\textwidth, height=.35\textwidth]{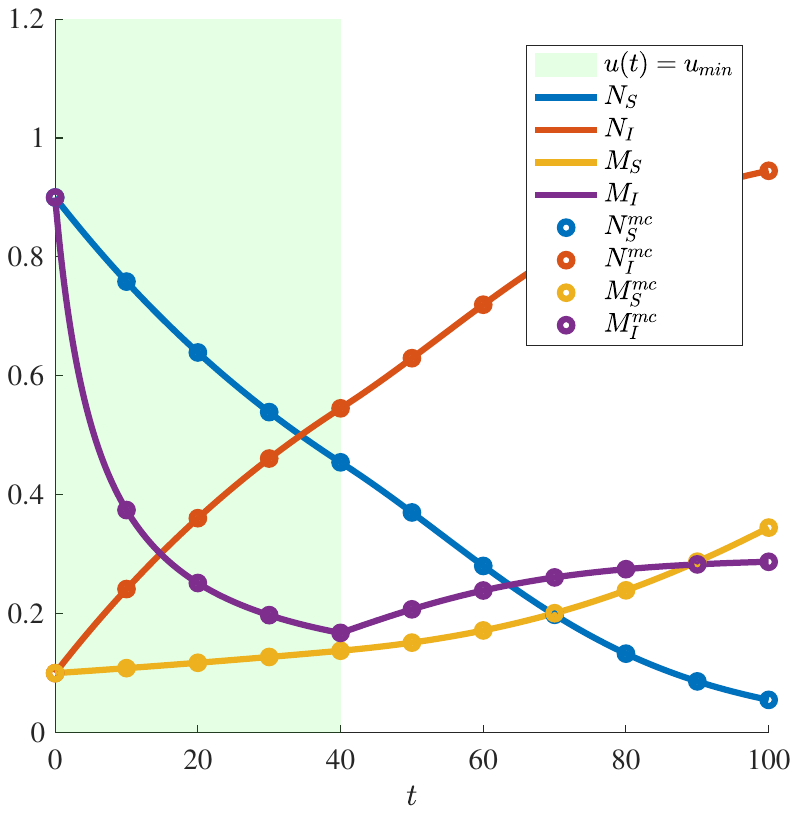}}
\subcaptionbox{\label{SIV5R5_beta_q11=0}}
{\includegraphics[trim={0cm 7cm 0 7cm},clip,width=.45\textwidth, height=.35\textwidth]{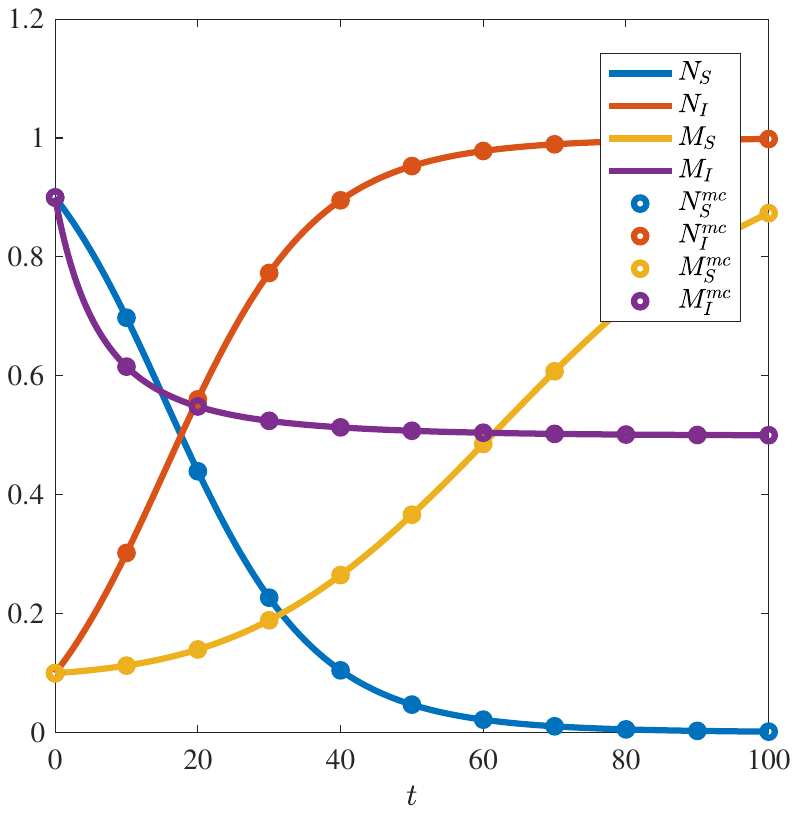}}
\caption{{ \bf Optimal solution extending Theorem 2 and the uncontrolled dynamics.} On the right the trajectories of the functions $N_S$, $N_I$, $M_S$, and $M_I$ without any control, obtained by prescribing $q$ defined via \eqref{q_2}. On the left, the optimal trajectories obtained by implementing the corresponding optimal control problem, extending the time horizon.}
\end{figure}

\end{document}